\documentclass[manuscript]{aastex}
\usepackage{graphicx}
\usepackage{rotating}
\renewcommand{\vec}[1]{\mbox{\boldmath$#1$}}
\newcommand{\kms}[1]{$#1$~km~s$^{-1}$}

\shorttitle{Emergence of highly sheared field and blowout eruptions}
\shortauthors{Lim et al.}

\begin{document}

\title{Observations of a Series of Flares and Associated Jet-like Eruptions Driven by the Emergence of Twisted Magnetic Fields}

\author{Eun-Kyung Lim\altaffilmark{1}, Vasyl Yurchyshyn\altaffilmark{1,2}, Sung-Hong Park\altaffilmark{3}, Sujin Kim\altaffilmark{1}, Kyung-Suk Cho\altaffilmark{1,4}, Pankaj Kumar\altaffilmark{1}, Jongchul Chae\altaffilmark{5}, Heesu Yang\altaffilmark{5}, Kyuhyoun Cho\altaffilmark{5}, Donguk Song\altaffilmark{5}, and Yeon-Han Kim\altaffilmark{1,2}}
\altaffiltext{1}{Korea Astronomy and Space Science Institute 776, Daedeokdae-ro, Yuseong-gu, Daejeon, Republic of Korea 305-348}
\altaffiltext{2}{Big Bear Solar Observatory, New Jersey
Institute of Technology, 40386 North Shore Lane, Big Bear City, CA
92314-9672, USA}
\altaffiltext{3}{Institute for Astronomy, Astrophysics, Space Applications and Remote Sensing (IAASARS), National Observatory of Athens, Penteli 15236, Greece}
\altaffiltext{4}{University of Science and Technology, Daejeon 305-348, Korea}
\altaffiltext{5}{Astronomy Program, Department of Physics and Astronomy, Seoul National University, Seoul 151-747, Korea}
\email{eklim@kasi.re.kr}

\begin{abstract}
{We studied temporal changes of morphological and magnetic properties of a succession of four confined flares followed by an eruptive flare using  the high-resolution New Solar Telescope (NST) operating at the Big Bear Solar Observatory (BBSO), Helioseismic and Magnetic Imager (HMI) magnetograms and Atmospheric Image Assembly (AIA) EUV images  provided by \emph{Solar Dynamics Observatory} (SDO). From the NST/H$\alpha$ and the SDO/AIA~304 \AA\ observations we found that each flare developed a jet structure that evolved in a manner similar to evolution of the blowout jet : 1) an inverted-Y shape jet appeared and drifted away from its initial position; 2) jets formed a curtain-like structure that consisted of many fine threads accompanied with subsequent brightenings near the footpoints of the fine threads; and finally 3) the jet showed a twisted structure visible near the flare maximum. Analysis of the HMI data showed that both the negative magnetic flux and the magnetic helicity have been gradually increasing in the positive polarity region indicating the continuous injection of magnetic twist before and during the series of flares. Based on these results, we suggest that the continuous emergence of twisted magnetic flux played an important role in producing a successive flares and developing a series of blowout jets.}
\end{abstract}


\section{INTRODUCTION}
A solar flare is one of the most remarkable eruptive events that vigorously releases  energy in a bursts-like manner and is often accompanied by a coronal mass ejection (CME). Due to their impulsive and energetic behavior, many observational and theoretical studies have been performed  to understand their energy source and the trigger mechanism \citep[references therein]{Asc02, Shi11}.

Flux emergence carrying magnetic twist from solar interior out into the solar atmosphere is known to {play an important role in providing magnetic energy for flares.} Many observational studies reported the manifestation of magnetic twist injection in active regions (ARs) prior to flares, such as sunspot rotation \citep{Ish98,Brown03,Yan07,Kum13} and existence of high magnetic shear along polarity inversion line \citep[PIL;][]{Brooks03,Lim10}. {the concept of magnetic helicity as a measure of magnetic twist in ARs , has been introduced and used to quantitatively describe this process.} Many observational \citep{Moo02,Nin04,Lab07,Park10} and theoretical \citep{Kus03,Yok03,Par09} studies showed the causal relationship between the injection of magnetic helicity and the onset of flares. These studies suggest that once the total accumulated amount of magnetic helicity in ARs exceeds the critical value necessary to break down the stability of magnetic environment, flares occur releasing magnetic energy out into  the corona and the heliosphere.

The flux emergence is also considered to be one of the trigger mechanisms for flare process . {\citet{Hey77} proposed a model in which solar flares occur in three consecutive phases  such as the pre-flare heating phase, the impulsive and flash phase, and the main phase in terms of emerging magnetic loops inside or near ARs. In the model, different types of flares, such as simple loop flares or two-ribbon flares occur based on the type of the pre-existing magnetic environment that interacts with the newly emerging field.} Magnetic flux carrying opposite polarity magnetic fields that emerges into (or near) a unipolar region is one of the well-known precursors of flares \citep{Zir74,Wan93,Fle01,Mas09,Wan14}. {When the emerging closed and twisted loops  reconnect with the overlying open fields, the interchange type of reconnection occurs thus enabling the twist to be transferred from the small-scale closed loops into the large scale coronal structures } {Similar magnetic configuration is also often adopted in 3D numerical simulations of magnetic reconnection along the quasi-separatrix layers \citep[QSL;][]{Dem97, Man97, Jan13, Man14}. In such models, continuous reconnection proceeds along the QSL where the field lines exchange their connectivity with the neighboring fields resulting in an apparent slippage of newly formed reconnected field lines \citep{Aul06, Mas09, Tor09}.

One noticeable observing signature of reconnection is a jet, i.e., plasma flows collimated along the reconnected field line. {Jet phenomena occurs on a variety of spatial and temporal scales, ranging from  granular size spicules \citep{Bec72} to chromospheric jets \citep{Shi07}, H$\alpha$ surges \citep{Roy73}, EUV jets \citep{Bru83}, and X-ray jets \citep{Shi92}.} Typical observed properties of a coronal jet are the inverted Y-shape and the apparent lateral drift away from the associated bright point. Such properties were successfully reproduced in numerical simulations of emerging closed field into ambient open field \citep{Yok96, Mor08}. Not only the coronal jets but also many chromospheric surges and macro-spicules may be explained by this magnetic configuration. {Recently \citet{Moo10} pointed out that one third of observed polar X-ray jets are of a non-standard type. Using X-ray jet data acquired by Hinode they classified these non-standard jets as ``blowout jets". They can be described as \emph{miniature versions of the blowout eruptions} similar to the major coronal mass ejections. According to \citet{Moo10}, such jets are non-standard in the sense that the base arch does not remain static but erupts accompanied with a brightening in its interior, and the jet's spire has an extra strand rooted near the chromospheric bright point. In order to understand such distinctive characteristics, \citet{Moo10} suggested a modified magnetic configuration that consists of a highly sheared core field of an emerging flux and an additional adjacent open field. Although the  blowout jet starts in the same way as the standard jet, due to the high degree of magnetic twist in the emerging field a more complex reconnection process, such as tether-cutting or breakout, can further develop in addition to the initial interchange reconnection. Then a blowout eruption of the sheared core field can be unleashed as a result.

{In the scenario of \citet{Moo10}, the main component of the blowout jet is the highly twisted emerging core field, so that a helical jet is expected to be observed. Similar ejecta-related twisted jets were also detected in previous studies and interpreted as being a result of reconnection between a twisted flux rope and the ambient field \citep{Shi92,Ohy98, Zha12, Nin15}.} {According to \citet{Moo10}, the removal of the outer layer of the emerging fields via reconnection is one possible mechanism that may lead to the rise of the sheared core and future blowout eruption. This breakout reconnection \citep{Ant98,Moo06,Kar12} may occur with or without the rising core field when the external current sheet forms. Other suggested possible mechanisms of the blowout eruption are MHD instabilities that may develop in the core field such as kink instability \citep{Tor04,Sri10,Kum12}. When the magnetic twist of the core field exceeds a certain threshold, for instance 1.25 turns in case of the uniformly twisted flux tube \citep{Hoo81}, the kink instability can be induced. Detailed high resolution multi-channel observations of the lower atmosphere are required in order to understand how the sheared core field begins to rise and leading to a blowout eruption.}

The high spatio-temporal resolution of the New Solar Telescope \citep[NST;][]{Goo10} benefits the studies of fine structure of the pre-flare process or the initial phase of the main flare, such as compact brightenings, jets, and  chromospheric fine threads. In order to understand the initiation and development of flares that we observed with the NST, we studied the time evolution of magnetic and morphological characteristics in association with flares.

\section{OBSERVATIONS AND DATA ANALYSIS}
{The NOAA AR 11515 first appeared at  the East limb in the southern hemisphere in 2012 June 27, and was located at (-110\arcsec, -330\arcsec) near the central meridian when the NST observation began on July 2 (Figure~\ref{ar_overview}). The adaptive optics \citep{Cao10a} aided NST observations that lasted for uninterrupted 5 hours from 16:28 UT to 21:38 UT under a good seeing condition.} Photospheric images were acquired every 15~sec using the broadband TiO filter imager \citep{Cao10b} working at $7057$~\AA\, with the pixel scale of $0\arcsec.0375$. Chromospheric data were obtained using a $0.25$~\AA\, bandpass Zeiss Lyot birefringent H$\alpha$ filter by shifting the central wavelength of the filter between the H$\alpha$ line center and the $-0.75$~\AA\ position from the line center. Different exposure times were used depending on the wavelength while the time cadence was close to 10~sec. The pixel scale of the H$\alpha$ data is $0\arcsec.0517$. The speckle reconstruction was applied to both TiO and H$\alpha$ data using the Kiepenheuer-Institute Speckle Interferometry Package \citep[KISIP;][]{Wog08} to obtain the diffraction limited resolution of  $0\arcsec.11$ at $7057$~\AA\,. {Sample TiO and H$\alpha$ images with the cropped field of view (FOV) are displayed in the lower panel of Figure~\ref{ar_overview}}.

We also used the data from the Atmospheric Image Assembly \citep[AIA;][]{Lem12} and the Helioseismic and Magnetic Imager \citep[HMI;][]{Sche12, Scho12} onboard the \emph{Solar Dynamics Observatory} \citep[SDO;][]{Pes12}. More specifically, the full-disk EUV images from AIA $171$~\AA\, and $304$~\AA\, channels with the pixel scale of $0\arcsec.6$ and the time cadence of $12$~sec were utilized to analyze the coronal structures, and the HMI full-disk longitudinal magnetograms with the pixel scale of $0\arcsec.5$ and the cadence of $45$~sec were used to understand the time evolution of magnetic flux.

The injection rate of magnetic helicity via photosphere in the AR was measured using HMI magnetograms and following the method described in \citet{Par05}. Firstly, the flux density of the magnetic helicity, $G_\theta (x, t)$ was calculated as:
\begin{equation}
\vec{G}_\theta (\vec{x},t) = -{{\vec{B}_n}\over{2\pi}} \int_{S_E} ({{\vec{x}-\vec{x}^\prime}\over{|\vec{x}-\vec{x}^\prime|^2}} \times [\vec{u}-\vec{u}^\prime])_n \vec{B}^\prime_n dS^\prime,
\end{equation}
where the subscript $n$ indicates the vertical component of the magnetic field $\vec{B}_n$, and $S_E$ is the entire photospheric surface of the AR, while \vec{u} is the velocity of the apparent horizontal motion of the magnetic footpoints obtained by the Differential Affine Velocity Estimator method \citep[DAVE;][]{Sch06}. Then the accumulated amount of injected magnetic helicity was determined by integrating $\vec{G}_{\theta}(\textit{\vec{x}},t)$ over the entire area the area and the studied time interval.

\section{RESULTS}
{As shown in Figure~\ref{ar_overview}, the NOAA AR 11515 has a complex magnetic field configuration. While its trailing part had only negative polarity fields, the leading part consisted of both positive and negative flux. This complexity of the leading part is also evident in the AIA~171\AA\, image. While the majority of large scale coronal loops were connecting the positive and negative fluxes of the leading and trailing parts, there also existed relatively smaller loops connecting the positive and negative elements within the leading part. Because of the complexity of the magnetic structures, this AR produced total of 11 flares, including 2 M-class flares, only on July 2, and 5 of them, including a M3.8 flare, were detected by the NST during the 5-hour observing run. All of these 5 flares occurred in the leading part, more precisely within the region outlined by the black square in the upper panel of Figure~\ref{ar_overview}. This AR continued to produce plenty of C and M-class flares on the following days until it turned over the west limb.}

{When we only consider the leading part, the TiO image in the lower panel shows that this region consisted of four positive polarity sunspots and one of them was outside of the FOV, and a number of pores with the negative polarity. We will refer to some of these sunspots and pores as  either ``P" or ``N" depending on their magnetic polarity i.e., positive or negative, respectively. If we focus on the negative flux region within the white box surrounded by sunspots P${1,2}$, and $3$ (Figure 1, lower left panel), we can see many granules of abnormally narrow shape some of which are indicated  by the black arrow. Most of them are elongated along the direction connecting the positive sunspot P1 and the negative pore N1. The dark threads seen in the H$\alpha$ image (white arrow, Figure 1, lower right panel) are also lying along the direction connecting P1 and N1. {Such dark, long threads observed in the H$\alpha$ image represent relatively horizontal magnetic fields in the chromosphere while more vertical fields are shown as short spike-like structures.} Both TiO and H$\alpha$ data indicate the existence of a horizontal field in the lower atmosphere connecting these two polarities.}

{For better visualization we cropped and zoomed-in this region and displayed it in Figure~\ref{tio_incl}. These elongated granules are co-spatial with the photospheric transverse magnetic field, whose total strength $\sqrt{\vec{B}_{x}^2 + \vec{B}_{y}^2} = 400$~G is represented by green contours in the figure. The development of such elongated granules is an observational evidence of emergence of horizontal magnetic fields through the photosphere \citep{Sch10a,Sch10b,Lim11}. The distribution of the inclination angle of the magnetic field vector obtained from the Milne-Eddington inversion of the HMI vector magnetogram (right panel) also indicates the existence of the horizontal magnetic field with the average inclination angle of about 103\degr. Moreover, the elongation of both negative and positive magnetic fragments shown in the HMI magnetogram in Figures~\ref{ar_overview} and \ref{fregion} also suggests that the newly emerging negative magnetic field is more horizontal closer to  the photosphere at this instance \citep{Che07,Che08,Ste11}.}

Figure~\ref{fregion} shows the evolution of the negative flux in the leading part. For the sake of clarity we name this region as the ``emerging flux region (EFR)". Although we are interested in the emergence of the negative flux, term ``EFR" refers to a larger FOV covering not only the emerging negative flux but also the adjacent positive flux to which it connects. We can see that small patches of negative and positive flux first appear south-west of the positive spot at around 12:00~UT on July 1 and they immediately began to diverge from each other as they kept emerging. While the positive polarity patches mostly remained near the periphery of the positive polarity sunspot, the negative polarity patches continuously migrated in the south-east direction toward another positive sunspot, P3. From the figure, we can also see that the negative flux elements as they emerge move along a path that curls clockwise around the positive spot P1 rather than moving in the radial direction away from the spot. This signature indicates that the emerging flux is far from being potential and it carries significant amount of negative magnetic helicity.

We measured the temporal change of the magnetic flux over the entire AR and within the EFR during two days starting from July 1. Figure~\ref{fplot} shows that both positive and negative flux showed gradual increase during the investigated time period. Thus, the total positive and negative flux of positive and negative polarity in the AR increased in absolute value by $46.7\times10^{20}$ Mx and $39.9\times10^{20}$ Mx, respectively. The $15$\% imbalance between two polarities may be due to various reason such as the different field inclinations over the polarities, flux leaving the FOV, or slow cancellation at one footpoint. Note that only the line-of-sight magnetic field is used  in this computation. The negative flux in the EFR (bottom panel) continued to increase until 21:00~UT {on July 2} by $13.4\times10^{20}$ Mx, which is 34\% of the total increase of the negative flux in the entire AR. The flux-time profile in the bottom panel shows a nearly linear growth within three time intervals with the increase rate being slightly different for each of those intervals. The averaged increase rate for the negative flux in EFR is $-0.33\times10^{20}$ Mx~hr$^{-1}$.

The peak times for all the flares that occurred in the EFR provided by the GOES-15 satellite are indicated by the vertical lines, with the dashed/solid lines representing C-class/M-class flares. Interestingly, the C-class flares have been following nearly every 2 hour up until an M5.6 flare has initiated at 10:51~UT on July 2. After a 6-hour long quiet period another series of C-class flares was observed with the flares occurring every 1 hour, which was again followed by an M3.8 flare at 20:07~UT. We speculate that such repeating pattern of flare occurrence is related to the persistent flux emergence in the EFR.

The overall progress of three C-class and one M-class flares was investigated in this study through the AIA~304~\AA\, data. The first C2.0 flare was excluded from the analysis since its initial phase was not covered by NST observations. Figure~\ref{aia304} shows the initial, main and the final phases of each of the C-class flares, while the M-class flare is represented by 4 images that well describe the initiation and the eruption process. The brightening and the morphological change of the loop system associated with the flares are better visible in AIA~304\AA\ than in AIA~171\AA\, indicating that the associated temperature is relatively low. The comparison of the selected images for each C-class flare revealed that they all have similar morphological properties, i.e., they are homologous. These flares start as a brightening near the negative flux N1 (blue contour) accompanied by dark and bright jets. Then the bright area is extended toward the positive flux P1 (yellow contour) with jets that are slightly drifting in the same direction. The early brightenings near N1 appear a few minutes before the main flare begins near P1. These homologous properties of recurring C-class flares suggest that this series of C-class flares was initiated by the same mechanism and they proceeded according to the same scenario.

{In addition, the images in the top panel of Figure~\ref{aia304} show the presence of a dark loop-like structure (white arrows in the top panel) with one foot anchored at P1 and the other near N1.} This {structure} was clearly seen during the C7.4 flare, then began to rise just after the C2.5 flare with complex brightenings developing underneath it. During the C2.5 flare, the brightening below the {dark structure} expanded slowly occupying the region between P1 and N1 (Figure~\ref{aia304}, at 19:47:20~UT). Starting at about 19:50~UT the {structure} became destabilized and began to expand followed by the brightening until 20:02~UT. Considering the location of this AR ($0\arcsec, -320\arcsec$), such expansion and motion in the  south-west direction can be regarded as the upward motion, in other words, gradual eruption . {It seems that the initiation of the rising motion of this {dark structure} was closely related to the underneath brightenings. However the details of this process was not well resolved from the AIA data.} The onset of the M3.8 flare was co-temporal with the eruption of the bright material that formed a transient helical shape (bottom panel in Figure~\ref{aia304}, also see the movie). {Although the sense of twist of the erupted helical structure can not be readily distinguished from the image due to both projection effect and overlapping jet, we could infer the sign of the twist based on the rotation of the helical structure during the eruption. Careful examination of the AIA~304\AA\ movie revealed that the erupted helical structure was rotating counterclockwise. This is consistent with the untwisting motion of the left-handed twist \citep{Kum14}. Moreover, the shape of dark and bright sub-threads of the rising {loop-like structure} ({indicated by an arrow at 19:58:32~UT}) also indicates that the loop-like structure itself has the left-handed twist}. The rotating helical structure of the erupting jet is a clear observational evidence of releasing magnetic twist through eruption.

The dynamics of the dark loop-like structure followed by bright material was traced along the slit represented by the white line in the figure. The corresponding space-time plot (Figure~\ref{xt_plot_aia304}) shows a well distinguished track of the dark loop-like structure (indicated by black arrows). It displays a slow rise for the first 12~minutes, and sudden acceleration for about 3~minutes afterwards. The linear fit to the track yielded the speed of \kms{11} during the slow rise, and \kms{113} during the acceleration phase. After about of 14~minutes, only a trace of the bright material can be distinguished in the plot yielding the speed of \kms{401}. Although the speed inferred from this space-time plot may be underestimated due to the projection effect, the inferred values of \kms{113} and \kms{401} are consistent with the rising speeds of prominences \citep{Mcc15} and coronal jets \citep{Shi96, Pat08}.

Activities related to the flares observed in the chromosphere were also investigated using the NST H$\alpha$ data. Due to the limited FOV, mostly early phase of each flare that occurred near N1 was recorded by the NST. Although the main flaring region near P1 was outside the NST FOV, the high resolution data captured the details of the pre-flare and initial phases of the flares that were not resolved with the AIA instruments. Among the three C-class flares that showed similar aspects of evolution, only the C3.7 flare is described in this paper since it showed the dynamics of the related fine structures most clearly.

Figure~\ref{Hal_c3.7} shows the successive H$\alpha -0.7~$\AA\, images covering the C3.7 flare. The green overplotted curve represents the PIL. At 17:24:38~UT, about 18 minutes before the flare onset, a tiny bright point appeared near the PIL (indicated by an arrow) and was later accompanied by a dark jet (17:33:54~UT). A few minutes later the dark jet split into two parts, one of which remained stationary while the other one slowly drifted toward the north-west as shown in 17:36:46~UT and subsequent images. Interestingly, due to bending of the drifting part, the overall shape of this jet became inverted Y (from the root to top of the jet). The inverted Y-shaped drifting jet is very similar to the coronal jets produced via magnetic reconnection between emerging closed fields and the ambient open field \citep{Shi92,Sav07,Sav09,Moo10}. Note that the bright point near N1 and the inverted Y-shape jet first appeared well before the main C3.7 flare. These observations strongly indicate that magnetic reconnection similar to that associated with a coronal jet took place at the chromospheric level before the onset of main C-class flare. \citep{Kim01,Shi07}.

{Subsequent images in Figure~\ref{Hal_c3.7} show that this H$\alpha$ jet that first appeared at the pre-flare phase and continued to evolve developing morphological properties of so-called blowout jet. As the leading strand of the inverted-Y jet drifted laterally, a curtain-like structure consisting of many fine threads formed brightenings near its footpoints (yellow curves; at 17:43:52~UT). These threads were not well distinguished in the AIA data, but only the outer edges were seen as two strands of the jet. The brightening area near the footpoints was expanding north-westward toward P1 as the leading strand drifted in the same direction. Except for the fine scale threads, similar properties such as double-strand jets and complex brightenings at the base arch were also reported by \citet{Moo10} from their observations on blowout jets. As pointed out by these authors, such properties cannot be explained by the standard jet model. {Moreover, the curtain-like threads with sequential brightenings near their footpoints gave  us an impression that a chain of reconnection events may have taken place and each of these threads is a plasma flow collimated along the reconnected magnetic field.} At 17:50:12~UT near the peak time of the C3.7 flare, the leading strand of the jet became much darker and displayed a twisted structure near its root (inside the dashed oval). As it was in the case of the erupting helical structure during the M3.8 flare, this H$\alpha$ twisted structure also showed both the left-handed twist of fine threads in the image and the counterclockwise rotation in the movie.}

We measured the apparent lateral motion of the observed jet along the slit normal to the jet axis (white line in Figure~\ref{Hal_c3.7}). As shown in Figure~\ref{xt_plot}, the leading strand of the jet began to drift at around t=10~minute (i.e., 17:34~UT), and then accelerated at around t=21~minute. The speed of the jet's apparent lateral motion obtained via a linear fit was found to be \kms{9.9} before t=21~minute, then increased to \kms{19.3}, which is comparable to the chromospheric Alfv\'{e}n speed. The main C3.7 flare began at 17:42~UT, and reached its maximum at 17:50~UT. That is, the enhancement of the jet's lateral drift is nearly co-temporal with the onset of the main flare. {Because of the development of the H$\alpha$ brightening and the rotating twisted structure near the root of the jet, the leading strand became too faint and complex to follow after t=25~minute.} The apparent lateral motion of the jet may be projection of real motion that is a combination of pure horizontal and vertical component. Note that the dark jet is detected in the blue-wing of the H$\alpha$ at $-0.7$~\AA\ from the line center, indicating that it has upward velocity component of at least \kms{30}.

We used the amount of magnetic helicity injected through the photosphere during two days as a proxy of magnetic energy accumulation in the AR. Since the EFR is the main source region for  the observed flares, the helicity injection within the EFR was also measured. In Figure~\ref{hel}, the accumulated amount of helicity injected  in the entire AR (black dashed curve) and in the EFR (black solid curve) is compared to the unsigned magnetic flux. As evident from the plot, the sign of the magnetic helicity in both the AR and the EFR is negative and is consistent with the sign of twist inferred from the erupting helical structure at the onset of the M-class flare (Figure~\ref{aia304}) and the twisted H$\alpha$ jet (Figure~\ref{Hal_c3.7}). The negative sense of twist is also consistent with the sign inferred from the trajectory of the negative footpoint of the emerging flux that moved whirling clockwise around P1 (Figure~\ref{fregion}). The total amount of magnetic helicity accumulated in the EFR during 2 days is $-4.75\times10^{42}$~Mx$^{2}$ which is large enough to produce two flare associated CMEs \citep{Dev00} even without considering the content of the magnetic helicity in the pre-existing coronal field. According to the LASCO CME catalogue \footnote{http://cdaw.gsfc.nasa.gov/CME\_list}, both M-class flares on July 2 (see Figure~\ref{fplot}) were related to CME eruptions. This analysis indicates that the newly emerging flux in the EFR carried enough of magnetic twist required to power two moderate sized eruptions.

\section{SUMMARY AND DISCUSSIONS}
Using 5-hour long NST and \emph{SDO} observations of an emerging flux region in NOAA ¬¡¬² 11515, we studied a series of four homologous C-class flares that recurred approximately every $1$ hour followed by an M-class flare with  an erupting helical structure. All the C-class flares showed similar morphological properties such as: 1) appearance of a relatively weak bright point between the emerging negative flux and the pre-existing positive flux; 2) development of an inverted Y-shape jet several minutes before the onset of the main flare; 3) splitting of the jet into two strands, with one drifting away and the other one remaining stationary; 4) appearance and expansion of a brightening near the footpoint of the drifting strand. The NST/H$\alpha$ data taken for the C3.7 flare showed that: 1) the jet had a curtain-like structure that consisted of many fine scale threads; 2) a small brightening formed near the footpoint of each thread; and 3) during the flare maximum a twisted structure is developed in the lower part of the visible jet.

The confined C-class flares accompanying jets developed in a manner reminiscent of the blowout jets described in \citet{Moo10}. {First, the brightening at the jet footpoint did not remain stationary near the initial location but expanded as some of the jet's strands drifted away. Second, a number of additional fine-scale threads appeared behind the drifting strands forming a curtain-like structure. The expanding brightening in higher resolution data turned out to be located at the footpoint of each fine thread. This curtain-like structure was also described in \citet{Moo10}, but its fine structure was not clearly resolved. Since these observed properties can not be explained by the well-known model of standard coronal jets, \citet{Moo10} proposed that a highly twisted core field exists in the base arch of the jet thus enabling the the jet to evolve in a way similar to the blowout CME eruption. Although the jets studied by \citet{Moo10} were mostly X-ray jets emanating from a coronal hole, the explanation can be also applied to our observations based on the morphological similarities and the configuration of the photospheric magnetic field. Moreover, measurements of the magnetic helicity injection rate in the flux emergence region strongly support the idea that the emerging magnetic field can indeed be described as a highly sheared core fields. }

The NST/H$\alpha$ data indicates that the magnetic reconnection at the initial phase of each C-class flare took place in the chromosphere or the lower transition region. Although the lateral extension of the emerging field was about $30$~Mm, it appears that this field was flattened near the surface as it emerged mainly because the strong overlying canopy fields may have prevented it from freely expanding into the atmosphere \citep{Lim13}. The elongated TiO granules, the elongated shape of complex negative and positive magnetic patches seen in HMI magnetograms, and large average inclination angle of $103$\degr\ all support this idea. \citet{Moo10} also argued that the reconnection relevant to blowout jets takes place at the lower height inside the base arch of emerging field.

{Our proposed interpretation of observed events is based on the NST/H$\alpha$ observations and is shown in Figure~\ref{cartoon}. This is similar to the one proposed by \citet{Moo10}, but slightly modified to reflect particularity of the observations. In the cartoon, pre-existing open field lines are represented by black curves and the emerging closed field lines are plotted with green curves. The undulating shape of the green lines signifies the magnetic twist carried out by the emerging fields. The left panel shows the initiation of the jet near the bright point during the pre-flare phase. The emerging closed fields connecting P1 and N1 reconnect with the ambient open fields rooted in P3 (marked by the X) as the negative flux N1 collides with the positive P3. Field lines before/after the reconnection are shown in gray/red. This picture explains the first and the second panel of Figure~\ref{Hal_c3.7}. It seems that reconnection between N1 and P3 continuously proceeded during our observations. The brightpoint with dark ejecta was always present in the AIA and H$\alpha$ FOV during all five flare events. We suggest that this continuous reconnection peeled-off the outer layer of the emerging field thus releasing the higher twisted core fields. The peel-off process can also explain the formation of the curtain-like structure consisting of fine threads shown in the H$\alpha$ data.}

{The right panel shows that the newly formed kinked orange field line drifts away from the reconnection site due to magnetic tension. The magnetic twist that the emerging fields carried out is transferred to open field lines via the interchange reconnection process as evidenced from the detected apparent rotating motion of the H$\alpha$ jet in Figure~\ref{Hal_c3.7}. Also, the schematic cartoon shows that sheared closed field slightly expanded and reconnected with overlying field from P2. We suggest that the process of removing of the overlying fields facilitates  the expansion of the core fields upward toward an unstable state. }

{As pointed out by \citet{Moo10}, the key component of the the magnetic configuration relevant to blowout jets is an emergence of highly sheared magnetic flux, which is the main source of magnetic energy required for the eruption. They suggested that as the flux emergence continues, the sheared core field will expand and evolve but remain in the equilibrium as long as the stabilizing force (in our case the overlying canopy fields) are strong enough to counter the expanding sheared fields. This equilibrium state can be broken by a reconnection process occurring either under the expanding flux (i.e., ``tether cutting") or above it (``the breakout"). In this case, the peeling-off observed here is a signature of the breakout process when the overlying fields were gradually removed from above and the stabilizing force holding down the emerging flux slowly weakened. Generally speaking the tether cutting reconnection could have proceeded along with the observed breakout \citep{Kli14}, however, the relatively small-scale of the erupted fields and enhanced chromospheric activity did not allow us to make any reliable inference. At this moment of evolution the sheared core fields may begin a slow rise and further expansion to adapt to the new magnetic configuration, while the continuous emergence further fuels the breakout reconnection. When the overlying stabilizing fields weaken below a critical level either torus or kink instability may set in which will cause the core fields to rapidly accelerate and erupt. This kind of speed profile was detected in our data. We did not detect in our data any signatures of loop writhing, which is the well-known observational sign of kink instability. The torus instability does not have any specific and unique signatures which would help us to easily identify it. However, the removal of the overlying fields, slow rise and expansion of the core fields, and absence of writhing all indicate an involvement of torus instability in the eruption process \citep{Aul10}. We thus interpret the series of flares reported here as being driven by continuous flux emergence accompanied by magnetic breakout and followed by an eruption triggered by torus instability.}

\acknowledgments Authors are grateful to the anonymous referee for constructive criticisms and comments that greatly improved the manuscript. BBSO operation is supported by NJIT, US NSF AGS-1250818 and NASA NNX13AG14G, and NST operation is partly supported by the Korea Astronomy and Space Science Institute and Seoul National University. The SDO data were (partly) provided by the Korea Data Center (KDC) for SDO in cooperation with NASA and SDO/HMI Team, which is operated by the Korea Astronomy and Space Science Institute (KASI). E.-K.L., K.-S.C., and S.K. are supported by the ``Planetary system research for space exploration" from KASI. K.-S.C. and S.P. acknowledge support by the U.S. Air Force Research Laboratory under agreement number FA 2386-14-1-4078. S.P. is also supported by the project ``SOLAR-4068" under the ``ARISTEIA II" Action. This work was conducted as part of the effort of NASA Living with a Star Focused Science Team ``Jets'' funded by NASA LWS NNX11AO73G grant. We thank ISSI for enabling interesting discussions. V.Y. also acknowledges support from AFOSR FA9550-12-0066 and NSF AGS-1146896 grants. The work of the SNU team is supported by the National Research Foundation of Korea (NRF - 2012 R1A2A1A 03670387).

\clearpage

\begin{figure}[tb]
\begin{center}
    \includegraphics[width=1\textwidth]{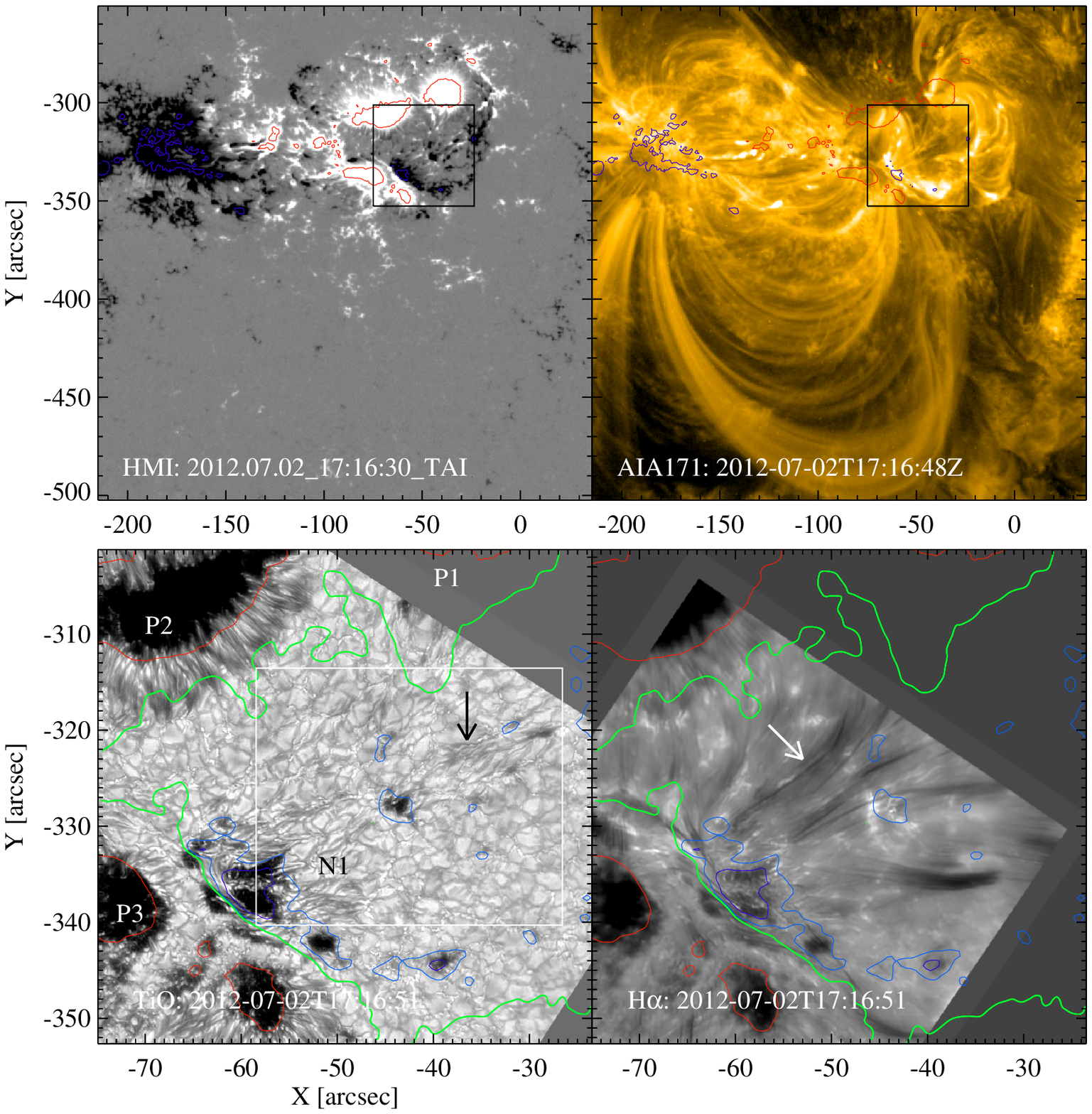}
    \caption{Active region overview just before the C3.7 flare onset. SDO/HMI (upper-left) and AIA~171~\AA\ (upper-right) images show the entire active region. NST/TiO (lower-left) and H$\alpha$ (lower-right) data have the same field of view (FOV) as the black squares in the top panels. Contours with purple, blue and red colors represent the longitudinal magnetic field strength of [$-1000, -500, 1000~$G], respectively. The polarity inversion line (PIL) is overplotted on the NST data as a green color curve. The positive sunspots are referred as a letter `P' and the negative as `N'. The cutout TiO image within the white box is enlarged in the Figure.~\ref{tio_incl}}\label{ar_overview}
\end{center}
\end{figure}

\begin{figure}[tb]
\begin{center}
    \includegraphics[width=1\textwidth]{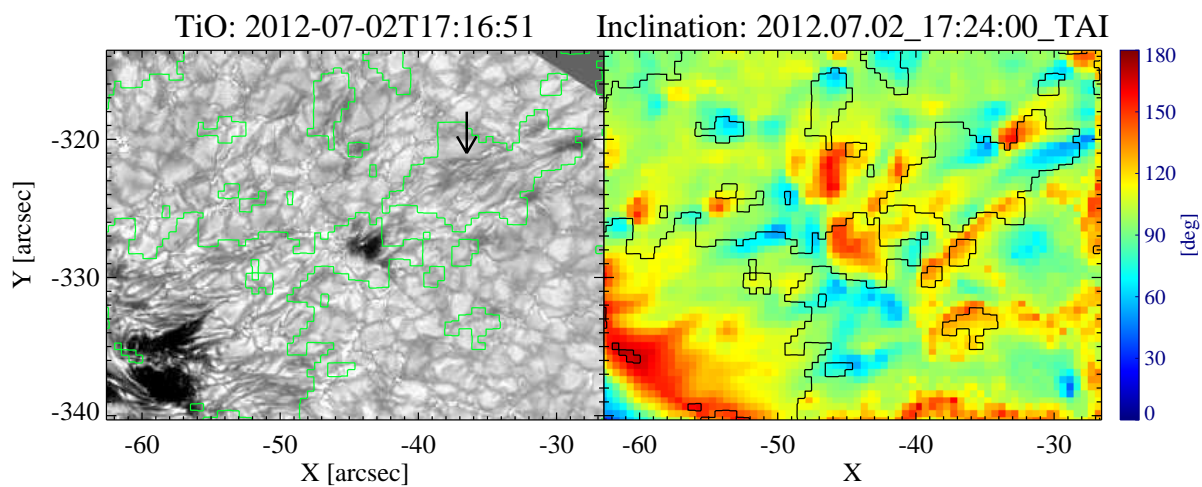}
    \caption{TiO image within the white box in the Figure.~\ref{ar_overview} (left) and the distribution of the inclination angle that was computed from the HMI vector magnetogram by applying the Milne-Eddington (ME) inversion (right). Both green and black contours in each panel represent the total transverse magnetic field strength, $\sqrt{ \vec{B}_{x}^2 + \vec{B}_{y}^2 } $, of 400~G. ME inverted HMI vector magnetograms are provided by the SDO. Some of elongated granules detected in the TiO images are pointed by the black arrow.}\label{tio_incl}
\end{center}
\end{figure}

\begin{figure}[tb]
\begin{center}
    \includegraphics[width=1\textwidth]{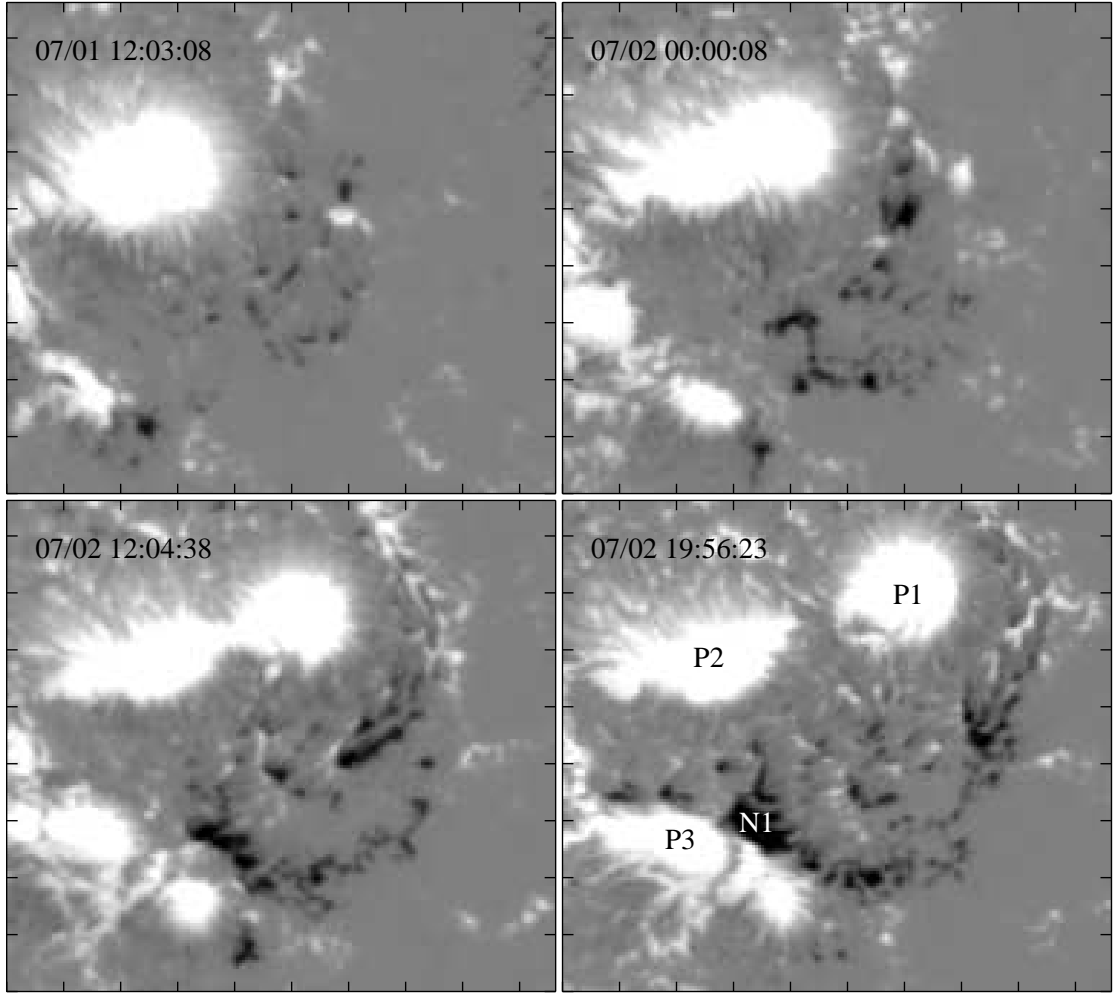}
    \caption{SDO/HMI longitudinal magnetograms showing the emergence of the negative flux in the leading part of the active region for two days. The interval between two tick marks corresponds to 10\arcsec.}\label{fregion}
\end{center}
\end{figure}

\begin{figure}[tb]
\begin{center}
    \includegraphics[width=0.8\textwidth]{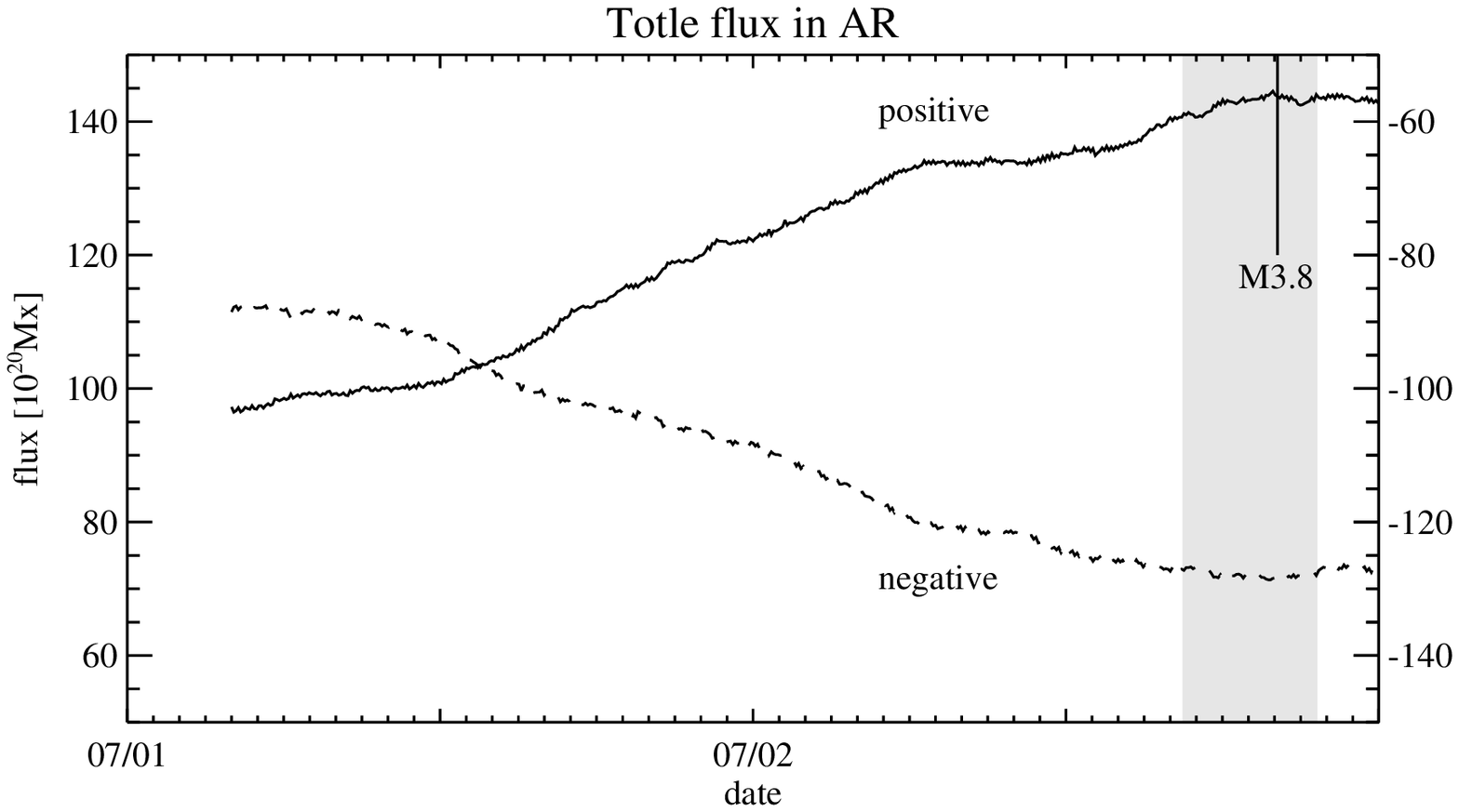}
    \includegraphics[width=0.8\textwidth]{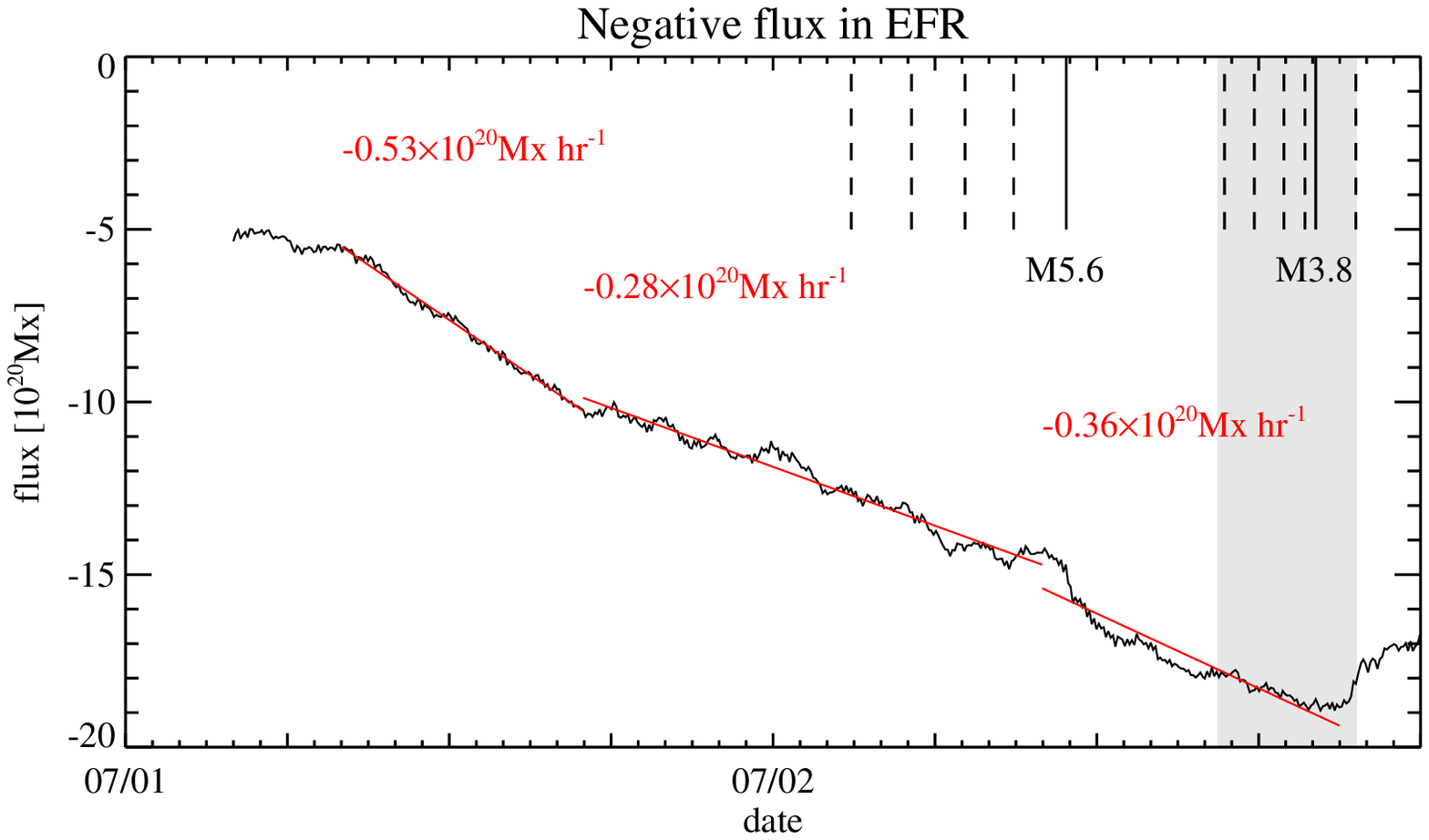}\\
    \caption{Temporal change of the total positive (top: solid curve) and negative (top: dashed curve) flux of the entire active region (top) and the negative flux within the EFR (bottom). Shaded rectangular area marks the NST observations time interval, and the dashed/solid vertical lines represent the maximum time of each C/M-class flares occurred in the EFR. The line segments in the bottom plot represent the linear fit of the negative flux injection rate. The averaged rate is estimated as $-0.33\times10^{20}$Mx~hr$^{-1}$.}\label{fplot}
\end{center}
\end{figure}

\begin{figure}[tb]
\begin{center}
    \includegraphics[width=1\textwidth]{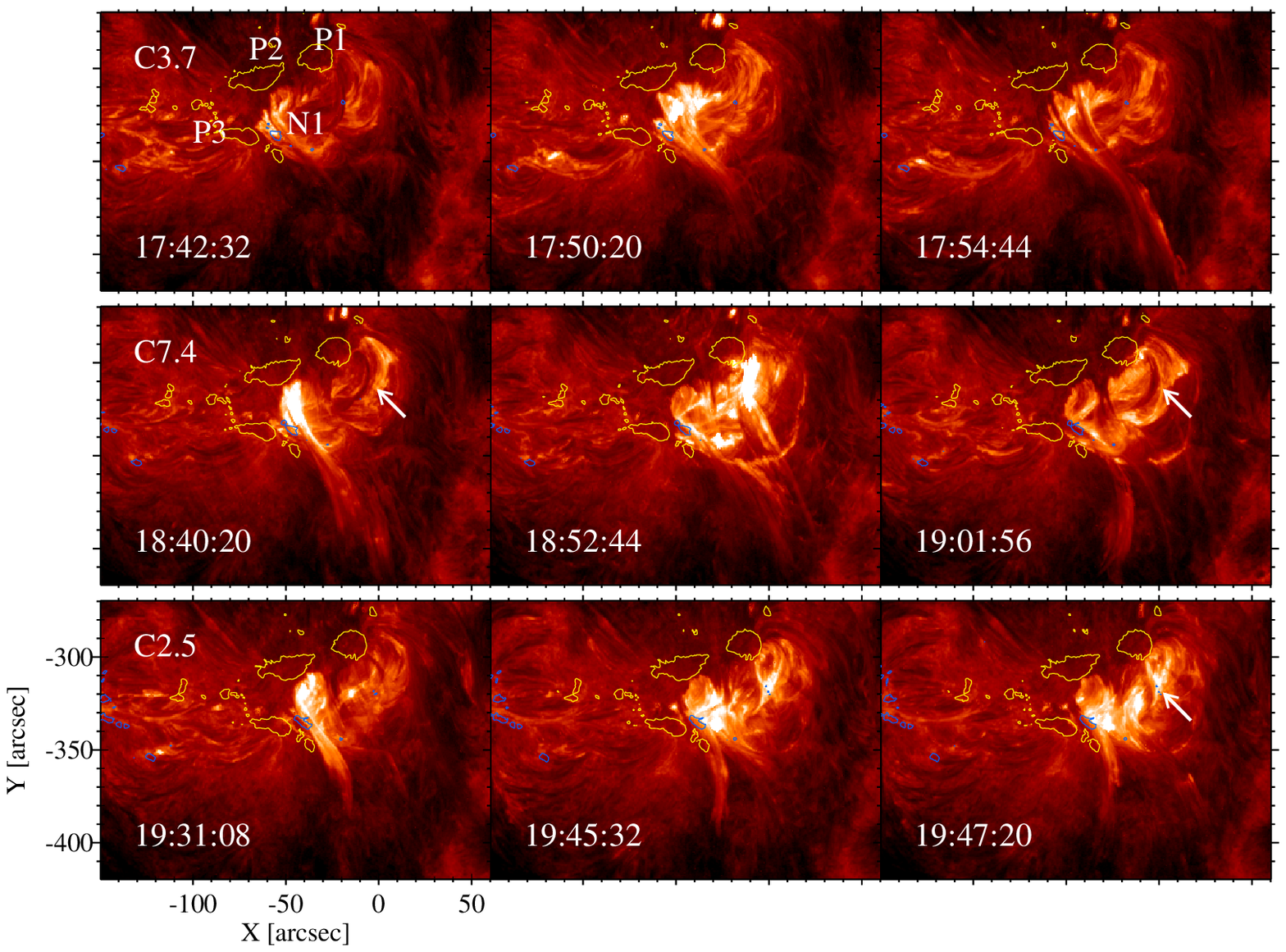}\\
    \includegraphics[width=1\textwidth]{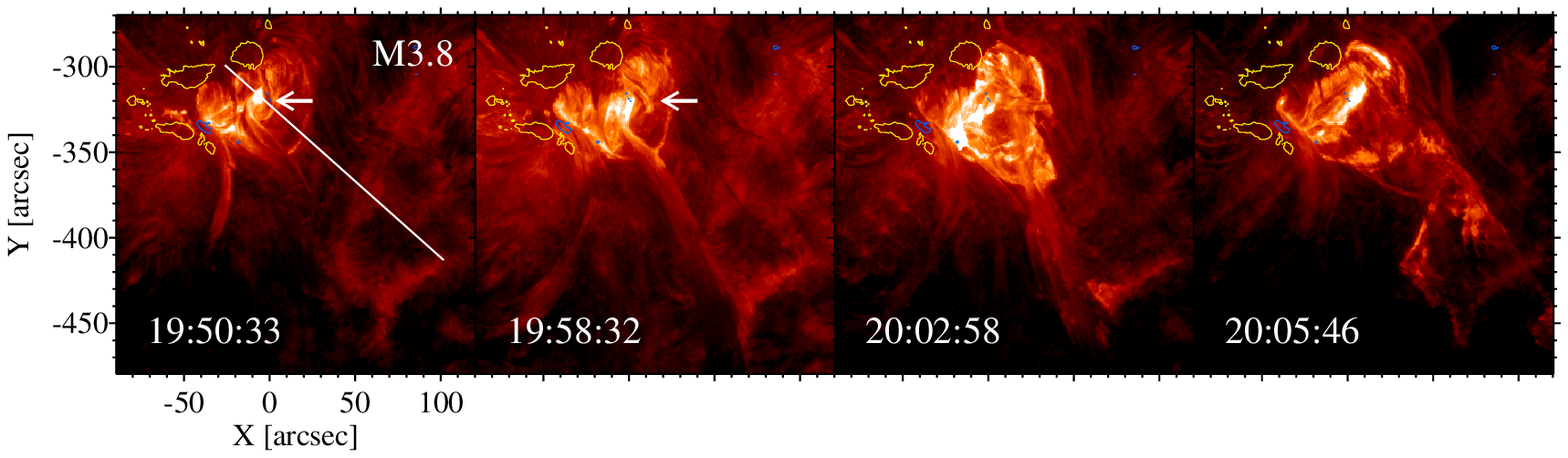}
    \caption{Top: AIA~304 images selected at the initial, main and the final phase of each C-class flare that occurred during the NST observations. Magnetic field contours are overplotted at levels of $\pm1000$~G with yellow (positive) and blue (negative) colors. Bottom: AIA~304 images showing eruption with the M3.8 flare. The white line indicates the slit along which the space-time plot was computed. White arrows in both top and bottom panels indicate dark, filament-like structure that erupted during the M3.8 flare. (An animation is available in the online journal.)}\label{aia304}
\end{center}
\end{figure}

\begin{figure}[tb]
\begin{center}
    \includegraphics[width=1\textwidth]{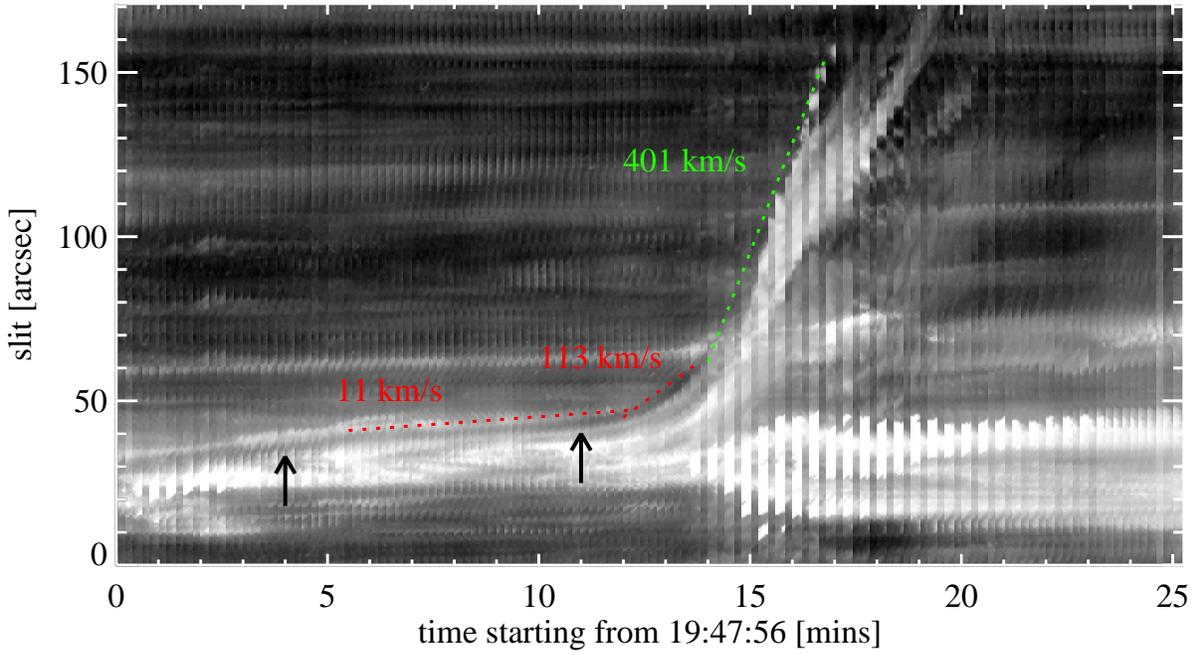}
    \caption{Space-time plot computed along the slit shown in Figure~\ref{aia304}. The track of the dark filament is pointed by black arrows, and the values of speed measured from the linear fit for each time intervals are indicated.}\label{xt_plot_aia304}
\end{center}
\end{figure}

\begin{figure}[tb]
\begin{center}
    \includegraphics[width=0.9\textwidth]{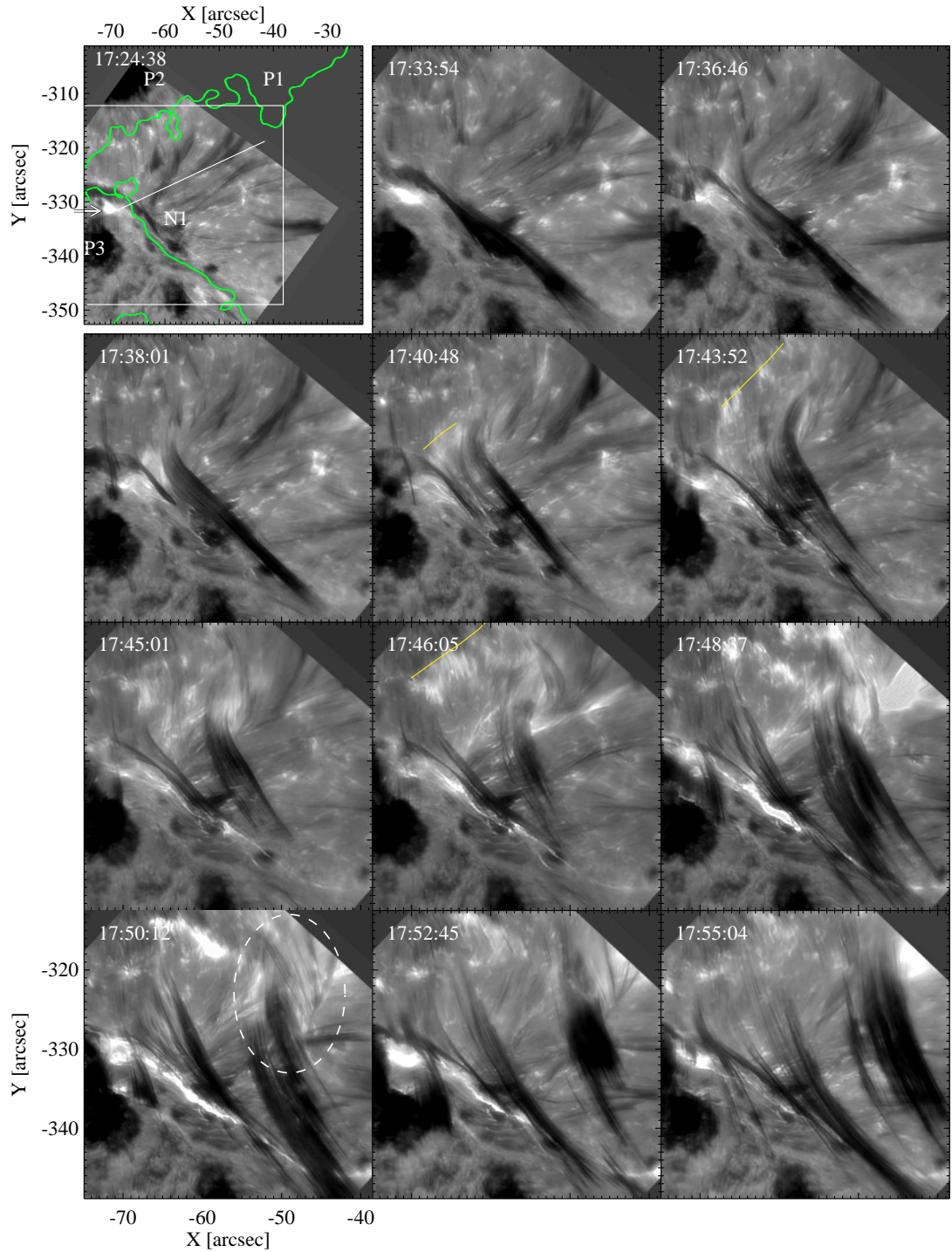}
    \caption{Time series of NST H$\alpha -0.7\,\AA$ images for C3.7 flare. A green curve in the first panel represents the PIL, and a white line represents the slit position along which the space-time plot was computed. Yellow curves indicate the position of sequential brightenings near the footpoints of curtain-like structure. (An animation is available in the online journal.)}\label{Hal_c3.7}
\end{center}
\end{figure}

\begin{figure}[tb]
\begin{center}
    \includegraphics[width=0.7\textwidth]{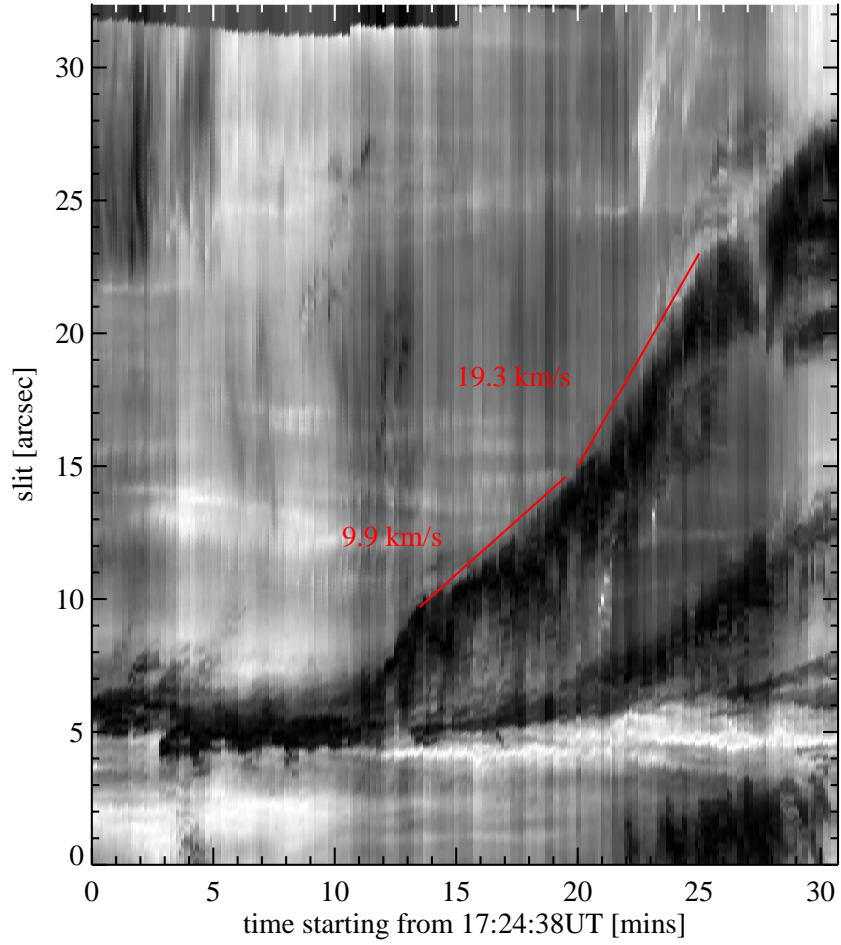}
    \caption{Space-time plot computed along the slit displayed in Figure~\ref{Hal_c3.7}.}\label{xt_plot}
\end{center}
\end{figure}

\begin{figure}[tb]
\begin{center}
    \includegraphics[width=1\textwidth]{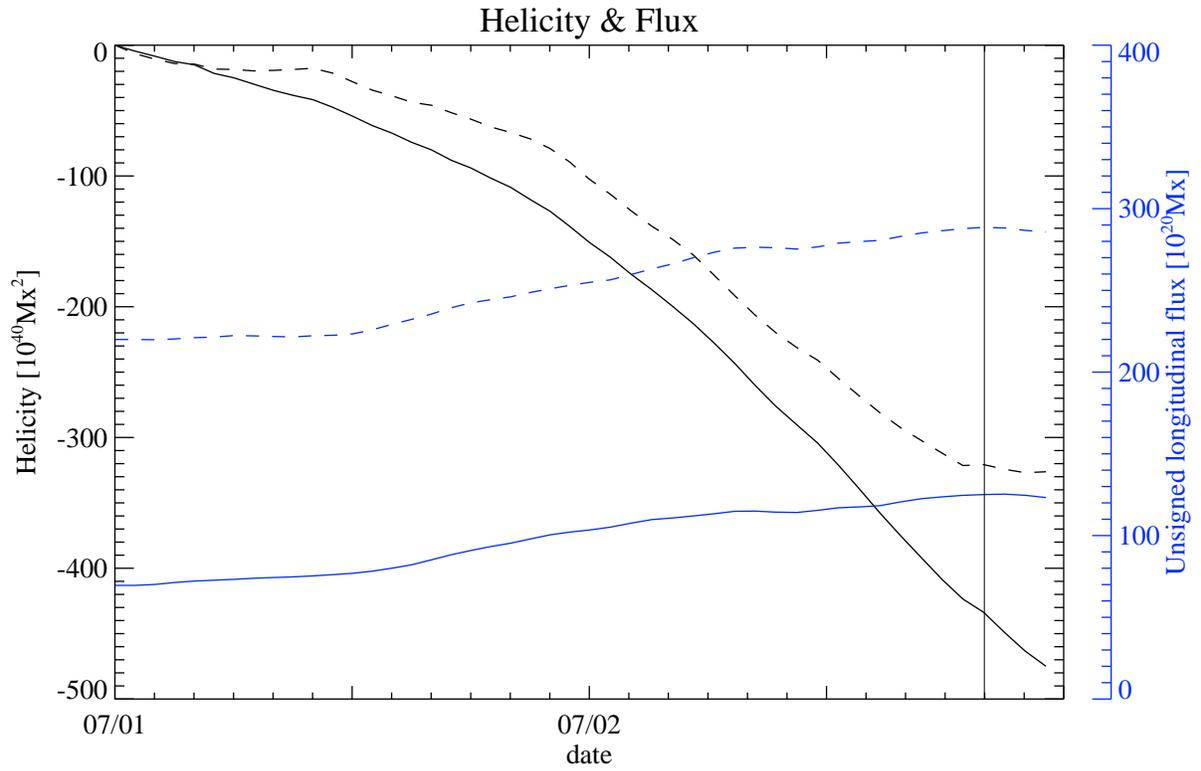}
    \caption{Temporal change of magnetic helicity injected in the entire active region/EFR (black dashed/solid curves) and the unsigned longitudinal flux in the active region/EFR (blue dashed/solid curves). 
    }\label{hel}
\end{center}
\end{figure}

\begin{figure}[tb]
\begin{center}
    \includegraphics[width=0.4\textwidth]{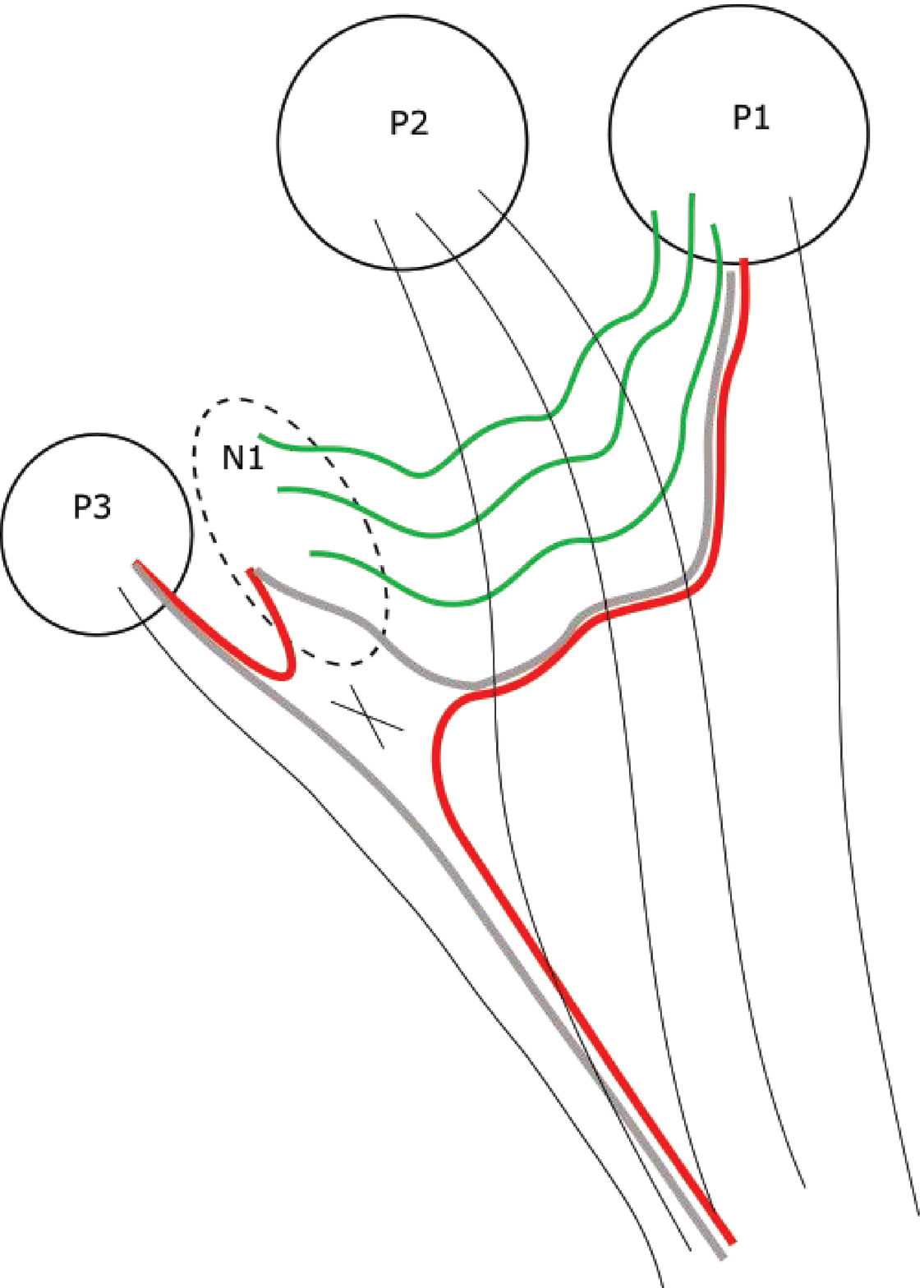}\,\,\,
    \includegraphics[width=0.4\textwidth]{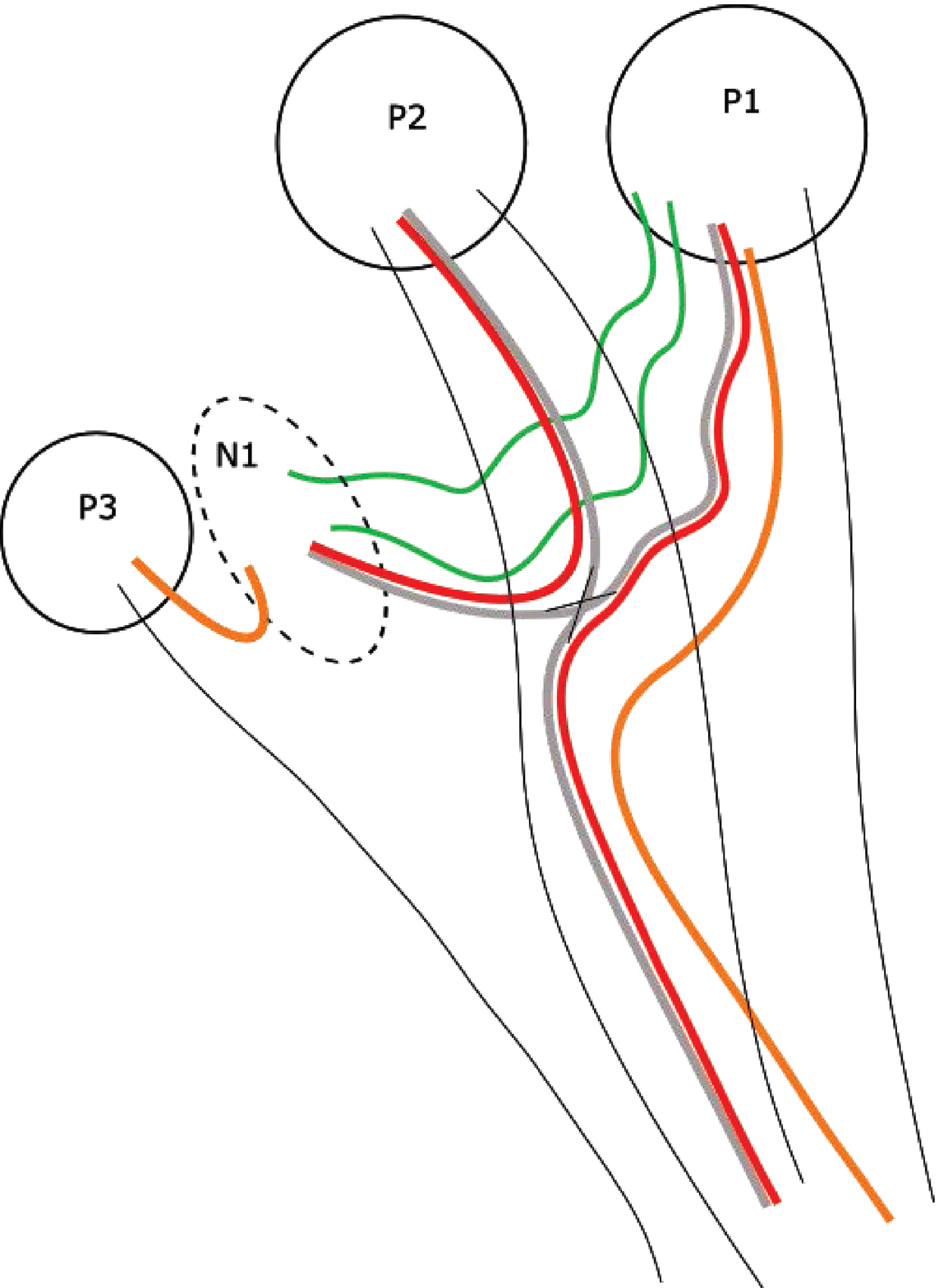}
    \caption{Schematic illustration of the magnetic field configuration in the EFR, with the same notations of `P' and `N' as in other figures and the text. Green lines represent emerging closed field and black lines represent open field. The undulating shape of green lines represents the magnetic twist of the closed field. Field lines before/after the reconnection (marked by the X) are expressed in gray/red, respectively.
    {Left: the initiation of the jet during the pre-flare phase. The emerging closed fields connecting P1 and N1 reconnect with the ambient open fields rooted in P3 as the negative flux N1 collides with the positive P3. Right: previously reconnected orange field line drifts away from the reconnection site due to magnetic tension. The magnetic twist that the emerging fields carried out is transferred to open field lines. Also, the sheared closed field slightly expanded and reconnected with overlying field from P2.}}\label{cartoon}
\end{center}
\end{figure}

\end{document}